# A study of the requirements of p-$^{11}$B fusion reactor by tokamak system code

Jianqing Cai[1,2], Huasheng Xie*[1,2], Yang Li[1,2], Michel Tuszewski[3], Hongbin Zhou[1,2] and Peipei Chen[1,2]

[1]Hebei Key Laboratory of Compact Fusion, Langfang 065001, China

[2]ENN Science and Technology Development Co., Ltd., Langfang 065001, China

[3]ENN Consultant, Riverside, CA 92506, USA

E-mail: *xiehuasheng@enn.cn, huashengxie@gmail.com

## Abstract

Most tokamak devices including ITER exploit the D-T reaction due to its high reactivity, but the wall loading caused by the associated 14MeV neutrons will limit the further development of fusion performance at high beta. To explore p-$^{11}$B fusion cycle, a tokamak system code is extended to incorporate the relativistic bremsstrahlung since the temperature of electrons is approaching the rest energy of electron. By choosing an optimum p-$^{11}$B mix and ion temperature, some representative sets of parameters of p-$^{11}$B tokamak reactor, whose fusion gain exceeds 1, have been found under the thermal wall loading limit and beta limit when synchrotron radiation loss is neglected. However, the fusion gain greatly decreases when the effect of synchrotron radiation loss is considered. Helium ash also plays an important role in the fusion performance, and we have found that the helium confinement time must be below the energy confinement time to keep the helium concentration ratio in an acceptable range.

## I. Introduction

A successful design of tokamak device is based on the good understanding of dependence of performance on key plasma parameters, such as density, temperature, magnetic field, current, etc. To study the dependence of these parameters, a tokamak system code (TSC) based on one-dimensional model has been developed, which considers a simple profile assumption.

The tokamak system code was firstly developed in the design of tokamak reactor since 1980s.[1,2] Stambaugh et al gave a more concise description of the system code to study the physical and engineer limits that restrict the design of fusion power reactor.[3,4] Then it has been extended by Costley et al to study the regime of steady-state reactors with high fusion gain, who found the fusion gain depends strongly on the fusion power and energy confinement, and weakly on the size of device, while the steady-state reactor is operating at fixed fractions of the density and beta limits.[5]

The tokamak system code has been used successfully on the deuterium-tritium (D-T) fusion reactor,[3-6] while the 14MeV-neutrons-induced damage and degradation would limit the useful lifetime of reactor component.[7] The neutron production in p-$^{11}$B fuel is lower than that in D-T and deuterium- deuterium (D-D) fuel by orders of magnitude. Furthermore, hydrogen and boron are abundant and fairly accessible on earth, while $^3$He might be mined and transported from the moon, which is one of the ingredients of deuterium-helium-3 (D-$^3$He) fuel. Therefore, there are sufficient reasons for us to pay more attention on proton-boron (p-$^{11}$B) fusion cycle. Since the core temperature of electrons is approaching the rest energy of electron, relativistic effect has to be considered in the extended tokamak system code to incorporate the relativistic bremsstrahlung. This extended tokamak system code is designed to include physics module, engineer module and economy module, the latter two of which are still under development and will be finished in the future work.

Of course, the ion temperature and $n_i\tau_E$ being high as 300keV and $10^{22}m^{-3}s$ for p-$^{11}$B fuel, is far beyond the current experiment condition. In this paper, we hypothesize the fuel ions could be heated up to the temperatures that p-$^{11}$B fusion reaction requires by certain method, mechanism of which is not the main point this paper discuss. Under this premise, we could further explore the p-$^{11}$B fusion cycle and study the requirements of p-$^{11}$B fusion reactor under some optimistic assumptions.

This paper is organized as follow: Sec 1 is the introduction of p-$^{11}$B fusion and the Tokamak System Code. Sec. 2 is the main equations used in this code. Sec. 3 is the results derived by the extended Tokamak System code. Sec. 4 is the conclusion and discussion.

## II. Equations of the tokamak system code

In this tokamak system code, a simple parabolic radial profile has been considered. Here, 0-dimensional model does not include the profile effects of density and temperature, whereas 1-dimensional model considers some certain density and temperature radial profiles.

The extended tokamak system code is developed from the models described in the paper by Stambaugh[3, 4], Petty[6] and Costley[5]. In terms of the fusion power of p-$^{11}$B fuel and the relativistic bremsstrahlung, the model has been revised accordingly, and the confinement enhancement factor H has been used here to measure the confining capacity needed for p-$^{11}$B reactor compared with the existing capabilities.

II.A. Geometry of tokamak

The geometry of tokamak size is described by parameters like aspect ratio $A$, major radius $R_0$, elongation $\kappa$ and triangularity $\delta$, from which other geometry parameters can be deduced:

The minor radius

$$a = R_0/A. \tag{1}$$

The elongation

$$\kappa = 0.9\kappa_{max} = 0.9(2.4 + 65\exp(-A/0.376)). \tag{2}$$

The plasma volume

$$V_p = \left(2\pi^2 \kappa A + \frac{16\pi k\delta}{3}\right)a^3. \tag{3}$$

The wall area

$$S_w = (4\pi^2 A\kappa^{0.65} - 4\kappa\delta)a^2. \tag{4}$$

Since low-aspect-ratio tokamak has the advantage of burning plasmas in a compact geometry at a lower cost than in conventional tokamak, we consider a low-aspect ratio tokamak device $A = 1.4$ and $\delta = 0.5$ as Figure 1 shows, which would prove to be a convenient choice later in the paper.

## II.B. Pressure and beta

In the p-$^{11}$B plasma, the density related parameters are as follows: the proton density $n_p$, the density of boron ion $n_B$, the ion density of p-B fuel $n_{i,pB}$, the electron density $n_e$ and the density of helium ion $n_{He}$, the profiles of which are described by the same parabolic form

$$n(x) = n_0(1 - x^2)^{S_n}, \tag{5}$$

where x=r/a is the normalized radial distance, the suffix zero in the subscripts means the core density and $S_n$ is the exponent of density profile. In this paper, densities are in units of $10^{20}$m$^{-3}$, and in what follows the units are generally m, s, T, MW, MA, keV.

Besides, the fractional ion densities are defined as $f_{pB} = n_p/n_B$, $f_B = n_B/n_{i,pB}$, $f_{He} = n_{He}/n_e$. From proportional relation, we have

$$f_{pB}f_B + f_B = 1.$$

From charge balance, we have

$$f_{pB}f_B n_{i,pB} + 5f_B n_{i,pB} + 2f_{He}n_e = n_e,$$

and core electron and ion density can be expressed as

$$n_{e0} = (f_{pB}f_B n_{i0,pB} + 5f_B n_{i0,pB})/(1 - 2f_{He}) \tag{6}$$

$$n_{i0} = n_{i0,pb} + n_{He0}. \tag{7}$$

The line-averaged density is

$$\overline{n_e} = \int_0^1 n_{e0}(1 - x^2)^2 dx. \tag{8}$$

The effective charge is

$$Z_{eff} = \frac{f_{pB}f_B n_{i,pB} + 25 f_B n_{i,pB} + 4 n_{He}}{n_e}. \tag{9}$$

And the average mass is

$$M = \frac{f_{pB}f_B + 5 f_B}{f_{pB}f_B + f_B}. \tag{10}$$

The temperature profiles are assumed to be the similar parabolic form

$$T(x) = T_0 (1 - x^2)^{S_T}, \tag{11}$$

and $S_T$ is the exponent of temperature profile, indicating this is one-dimensional model. The cross-section is considered to be elliptic in the integral computation, and triangularity is only used while calculating the volume and area. Refer to the previous papers,[5, 6] the density profile has taken $S_n$=0.5 for a broad H-mode profile with a pedestal, and the temperature profile was assumed steeper, $S_T$=1.5, which is shown in Figure 1.

The profile effect has been considered in the calculation of triple product to compare with the result of 0-dimensional mode, which is shown in Figure 2. Figure 2 illustrates that, Lawson criterion of D-T, D-$^3$He and D-D fusion reactor could meet at the same density of fuel ions, and the same ion and electron temperature whether in 0-dimensional model or in 1-dimensional model. But in the p-$^{11}$B fusion reactor, the optimum mix of proton and Boron is not 1:1, and the ion and electron temperature must also be different while the profile effect has been considered. Here we define the ratio of ion to electron temperature as

$$\frac{T_i}{T_e} = f_T. \tag{12}$$

The optimum mix of proton and Boron and the critical ratio of ion to electron temperature will be discussed in more details in Section 3. The toroidal beta, normalized beta and poloidal beta is given by

$$\beta_T = \frac{(0.04 n_{i0} T_{i0} + 0.04 n_{e0} T_{e0})}{(1+S_n+S_T) B_{T0}^2} * 100 \tag{13}$$

$$\beta_N = \frac{a \beta_T B_{T0}}{I_p} \tag{14}$$

$$\beta_p = \frac{25}{\left(\frac{\beta_T}{100}\right)} \left(\frac{1+k^2}{2}\right) \left(\frac{\beta_N}{100}\right)^2, \tag{15}$$

where $I_p$ is plasma current and $B_{T0}$ is core magnetic field strength.

## II.C. Fusion power

In p-$^{11}$B fuel, the fusion power element is written as

$$dP_{fus} = (1.6 \times 10^{15}) * 8.7 * n_p n_B (\overline{\sigma v}) dV,$$

in which the fusion reactivity is used from the recent evaluation by Sikora and Weller.[8] Using $x = r/a$, the fusion power can be represented as

$$P_{fus} = (1.6 \times 10^{15}) * E_{pB} * n_{p0} n_{B0} T_{i0}^2 * 2 V_p \int_0^1 \frac{(\overline{\sigma v})}{T_{i0}^2} (1-x^2)^{2S_n} x dx.$$

$\overline{\sigma v}$ is determind by x, since the fusion reactivity is a function of ion temperature, and ion temperature is a function of $x$. We use $\phi$ to replace with fusion reactivity integral

$$\phi = \int_0^1 \frac{(\overline{\sigma v})}{T_{i0}^2} (1-x^2)^{2S_n} x dx, \tag{16}$$

and once the core ion temperature and the profile exponent has been given, the fusion reactivity integral is determined. The core ion density of p-B fuel can be represented as

$$n_{i0,pB} = T_{i0}^{-1} (f_{pB} f_B^2)^{-1} \left(\frac{P_{fus}}{27.84 \times 10^{15} V_p \phi}\right)^{0.5}. \tag{17}$$

The fusion gain equals fusion power divided by auxiliary heating power

$$Q = \frac{P_{fus}}{P_{aux}}. \tag{18}$$

The condition of Q=1, when the fusion power is equal to the auxiliary heating power, is referred to as scientific breakeven. Since this paper is mostly focus on physics module, the engineering breakeven and economic breakeven are not discussed here.

## II.D. Plasma current

In p-[11]B reactor, an empirical current driven efficiency formula derived in tokamak has been used,[9] and the driven current can be calculated by

$$I_{cd} = \frac{0.031 \zeta_{cd} P_{cd} T_{e0}}{R_0 n_{e0}}. \tag{19}$$

$\zeta_{cd}$ is the dimensionless current drive efficiency, which is taken as 0.2 here. The safety factor can be calculated from

$$q_{eng} = \frac{5 B_{T0} a^2 k}{R_0 I_p}. \tag{20}$$

In addition to the driven current, the rest is the bootstrap current

$$I_{cd} = (1 - f_{bs}) Ip, \tag{21}$$

where $f_{bs}$ is the bootstrap current fraction. From Andrade and Ludwig,[10] the bootstrap current fraction $f_{bs}$ can be deduced from

$$f_{bs} = \frac{5 C_{bs}^* A^{0.5} C_p \beta_N q_{eng} (R_0/R_m)}{100 li^{1.2}}, \tag{22}$$

where $C_{bs}^* = 0.1558 \pm 0.0005$, $C_p = (1 + S_n + S_T)$, internal inductance $l_i$ has been taken as 0.5, and $\frac{R_0}{R_m} = 0.8$. We define a constant $const = \frac{5 C_{bs}^* C_p \left(\frac{R_0}{R_m}\right)}{li^{1.2}} = 7.16 C_p$, then substitute the equation of $\beta_N$ and $q_{eng}$ into equation (22), after which the bootstrap current fraction could be expressed as

$$f_{bs} = \frac{const A^{0.5}}{100} \frac{a B_{T0} \beta_t}{I_p} \left( \frac{5 B_{T0} a^2 k}{R_0 I_p} \right) = \frac{const F}{I_p^2}, \tag{23}$$

where $F = A^{0.5} \frac{(a^3 B_{T0}^2 \beta_T k / R_0)}{100}$. The plasma current and bootstrap current can be deduced by equation (21) and (23), and the solution is

$$I_p = \frac{I_{cd}}{2} + \frac{1}{2}(I_{cd}^2 + 4 const F)^{0.5}. \tag{24}$$

Hence the bootstrap current fraction could also be deduced. The Greenwald density limit is expressed as

$$n_{GW} = \frac{I_p}{\pi a^2}. \tag{25}$$

## II.E. Radiation

One of the main difficulties for p-$^{11}$B fuel is that the bremsstrahlung radiation power loss is even higher than fusion power without synchrotron radiation. In the non-relativistic case, the bremsstrahlung radiation power per unit volume is given by

$$\frac{dP_{brem}}{dV} = 0.00534 Z_{eff} n_e^2 T_e^{1/2}. \tag{26}$$

In the p-$^{11}$B reaction, the relativistic electrons have to be considered since electron temperature is approaching the rest energy of electron. The relativistic electron-ion bremsstrahlung power per unit volume is given by

$$\frac{dP_{brem}^{ei}}{dV} = n_e^2 Z_{eff} \alpha r_e^2 m_e c^3 \begin{cases} \frac{32}{3} \sqrt{\frac{2t}{\pi}} (1 + 1.78 t^{1.34}) & \alpha^2 \leq t < 1 \\ 12t \left( \ln(2\eta_E t + 0.42) + \frac{3}{2} \right) & t > 1 \end{cases}, \tag{27}$$

where classical electron radius $r_e = 2.818 \times 10^{-15} m$, fine structure constant $\alpha = 1/137$, electron mass $m_e = 9.11 \times 10^{-31} kg$, light speed $c = 2.998 \times 10^8 \, m/s$, elementary charge $e = 1.6 \times 10^{-19} C$, the normalized electron temperature with respect to the rest energy of electron $t = \frac{T_e}{m_e c^2}$ and $\eta_E \approx 0.5616$. [11, 12]

At this temperature, the bremsstrahlung radiation caused by the electron-electron scattering is comparable to that caused by the electron-ion scattering. And the relativistic electron-electron bremsstrahlung power per unit volume is[11, 12]

$$\frac{dP_{brem}^{ee}}{dV} = n_e^2 \alpha r_e^2 m_e c^3 \begin{cases} \frac{20 t^{1.5}}{9\sqrt{\pi}} (44 - 3\pi^2)(1 + 1.1t + t^2 - 1.25t^{2.5}) & \alpha^2 \leq t < 1 \\ \left[ 24t \left( \ln\left(2\eta_E t + \frac{5}{4}\right) \right) \right] & t > 1 \end{cases}.$$

(28)

Hence, the total bremsstrahlung power is

$$P_{brem} = 2V_p \int_0^1 \left( \frac{dP_{brem}^{ei}}{dV} + \frac{dP_{brem}^{ee}}{dV} \right) x \, dx. \tag{29}$$

In addition to bremsstrahlung radiation, synchrotron radiation is one of the important energy loss mechanisms in magnetically confined p-$^{11}$B plasmas, whose critical role has been discussed in the previous work,[13, 14] showing the p-$^{11}$B fuel can't generate net power in a magnetic confinement device with the synchrotron radiation loss. The synchrotron radiation loss could be obtained from

$$P_{cycl} = 4.14 \times 10^{-7} n_{eff}^{0.5} T_{eff}^{2.5} B_{T0}^{2.5} (1 - R_w)^{0.5} a_{eff}^{-0.5} \left( 1 + \frac{2.5 T_{eff}}{511} \right) V,$$

(30)

where $n_{eff} = n_{e0}/(1 + S_n)$ is the volume averaged density, $T_{eff} = \int_0^1 T_e(x) \, dx$ is the effective electron temperature, $R_w$ is the wall reflectivity, $a_{eff} = a\kappa^{0.5}$ is the effective minor radius and $V = 2\pi R_0 \pi a_{eff}^2$ is the approximation of the plasma volume.[15, 16] In this paper, the cases of considering the effect of synchrotron radiation loss and neglecting synchrotron radiation loss would both been discussed.

## II.F. Confinement

The confinement time could be obtained by

$$\tau_E = \frac{(0.024 n_{i0} T_{i0} + 0.025 n_{e0} T_{e0}) V_p}{(1+S_n+S_T)(P_{fus}+p_{aux}-p_{brem})}, \qquad (31)$$

and a confinement enhancement factor H is used here to compared the confinement time needed for p-$^{11}$B fuel compared with the confinement time predicted by ITER98y2 High-confinement mode (H-mode) scaling[17]

$$\tau_{98,y2} = 0.145 \frac{I_P^{0.93} R_0^{1.39} a^{0.58} k^{0.78} \overline{n_e}^{0.41} B_{T0}^{0.15} M^{0.19}}{(P_{fus}+P_{aux})^{0.69}} \qquad (32)$$

$$H = \frac{\tau_E}{\tau_{98,y2}}. \qquad (33)$$

## III. Results

In D-T, D-$^3$He and D-D fusion reactor, one of the main goals is to study the dependence of the fusion performance especially fusion gain Q on which plasma and devices parameters. But in p-$^{11}$B fusion reactor, a lot of optimistic assumptions and optimum parameters are needed to consider firstly to achieve Q=1, which is described by a simple logic diagram as shown in Figure 3.

In p-$^{11}$B fusion reactor, the hydrogen-boron mix needs to be carefully evaluated firstly to maximize the ratio of fusion power to radiation power. The range of ion temperature and $n_i \tau_E$ could be confirmed by the breakeven condition. Besides, the constraint of beta and thermal wall loading are also important, which are related with the magnetic field limit and ion density limit, respectively. Combined with the IPB98y2 scaling of energy confinement time, the effect of H factor and major radius on fusion performance can be studied.

### III.A. The concentration ratio of hydrogen to boron

The ratio of fusion power to radiation power $\frac{P_{fus}}{P_{brem}}$ is a function of ion temperature, electron temperature and the concentration ratio of hydrogen to boron $f_{pB}$. To obtain the optimum value of $f_{pB}$ which maximizes equation (34), the ratio of ion to electron temperature $f_T$ has been extracted, and the extremum problem has been simplified to maximizing $F(f_{pB}, T_e)$ by choosing the optimum value of $f_{pB}$.

$$\frac{P_{fus}}{P_{brem}} = F(f_{pB}, T_i, T_e) = f_T^2 F(f_{pB}, T_e). \tag{34}$$

A color map of $\frac{P_{fus}}{P_{brem}}$ versus electron temperature and $f_{pB}$ is shown in figure 4. The optimal value of ratio $f_{pB}$ =9:1, which maximizes $\frac{P_{fus}}{P_{brem}}$ with fixed electron temperature. Nevertheless, the maximum value of $\frac{P_{fus}}{P_{brem}}$ obtained in this simulation is still below 0.5, which indicates that higher ratio of ion to electron temperature is needed in order to obtain higher $\frac{P_{fus}}{P_{brem}}$.

### III.B. The ratio of ion to electron temperature

One of the main difficulties of p-$^{11}$B fuel is that the bremsstrahlung radiation power is much higher than the fusion power produced by fuel ions of p-$^{11}$B, as shown in figure 4, with equal ion and electron kinetic temperature

$$P_{fus} + P_{aux} = 2P_{fus} > P_{brem} + \frac{W_{dia}}{\tau_E}.$$

To study the dependence of Lawson criterion on $T_i/T_e$, we scan the $T_i/T_e$ ratio, and plot the minimum value of $n_i \tau_E$ and the corresponding ion temperature to meet fusion gain Q=1 as figure 5(a) shows, which indicates when $\frac{T_i}{T_e} < 1.12$, there are no positive values of $n_i\tau_E$, and to obtain a fusion gain greater than 1, the ion temperature is at least 1.12 times the electron temperature. In the later paper, the value of $T_i/T_e$ is assumed to be 2.5 in order to get a higher gain fusion while the value of $n_i\tau_E$ could be as low as possible.

### III.C. The constraint of thermal wall loading

The minimum value of $n_i\tau_E$ meeting the Lawson criterion of p-$^{11}$B fusion with $T_i/T_e$ =2.5 is $2.3\times10^{21} m^{-3}s$, which is still much higher than the current experiment parameters. To make a further study on the parameters of p-$^{11}$B fusion devices, a higher confinement enhancement factor H is assumed, not limited by the IPB98y2 energy scaling.

The p-$^{11}$B fusion reactor may be not limited by neutron irradiation, but thermal loading due to the significantly high bremsstrahlung radiation power at the ion temperature of 300-500keV would restrict the regime of ion density. In this paper, we take the maximum value of the thermal wall loading caused by radiation as 10MW/m$^2$, which is the maximum tolerable steady-state perpendicular power flux density onto the ITER divertor plate.[18] This wall-loading limit might further increase within the near-term development of materials technology.

If we assume all radiation power must be brought out through the first wall, we have the constraint of thermal wall loading

$$\phi_w = \frac{P_{brem}}{S_w} \leq 10,$$

where $\phi_w$ is the wall loading limit. The thermal wall loading is a function of the ion temperature, ion density and major radius, and for a given ion temperature, a larger major radius would restrict the increase of density.

In figure 6, the magenta and blue line are the bremsstrahlung radiation power curve corresponding to wall loading limit with $R_0$=2m and $R_0$=3m, which is also a limit of ion density for a given ion temperature. The parabola describes the value of $n_i \tau_E$ versus ion temperature satisfying fusion gain Q=1 while $\frac{T_i}{T_e} = 2.5$, where the minimum value of $n_i \tau_E = 2.3 \times 10^{21} m^{-3} s$ can be found at $T_i = 380 keV$. Since the fusion power density is directly related to the economic benefits, one approach to achieving economic benefits maximum is to operate at marginal ion density, which is shown by the circles in Figure 6.

## III.D. The constraint of Beta

A strong magnetic field is required to confine p-$^{11}$B plasma with extremely high pressure, which might exceed the engineer constraints greatly. In this paper, the on-axis magnetic field is designed below 20T considering foreseeable development of technology. The normalized beta $\beta_N$, which is related with economic benefits and plasma stability, is designed to be the maximum allowed for stability to confine the high-pressure p-$^{11}$B plasma.

To study the dependence of magnetic field on $\beta_N$, we combine equation (13) and equation (14), and substitute the value $\frac{T_i}{T_e} = 2.5, \frac{n_e}{n_i} = 1.4$ derived before into it, and obtain the relation of plasma current, magnetic field and $\beta_N$

$$\frac{2.08 n_{i0} T_{i0} a}{\beta_N} = I_p B_t.$$

Recalling the safety factor limit $q_{eng} > 2$, we have

$$B_{t0}^2 > \frac{4.16 n_{i0} T_{i0} R_0}{5 a \kappa \beta_N}.$$

To minimize magnetic field at fixed fusion power density and tokamak geometry, the normalized beta has been set to be the maximum $\beta_N \sim \frac{9}{A}$.[4, 5] In Table I $n_{i0} = 6\times 10^{20} m^{-3}, T_{i0} = 380 keV$ and $R_0 = 3m$ is assumed, the minimum value of magnetic field needed in p-$^{11}$B fuel is 10.7T for a low-aspect-ratio tokamak with A=1.4, while it turns out to be 23.2T for a conventional tokamak with A=2.4.

### III.E.    The size effects and H effects on fusion gain

As discussed above, the optimal value of $\frac{n_p}{n_B} = \frac{9}{1}, \frac{T_i}{T_e} = 2.5, T_i = 380 keV$ have been found, and two constraint conditions $\phi_w = 10 MW m^{-2}$ and $\beta_N = 6.4$ have been confirmed.

To study the dependence of fusion gain on H and major radius R under the input and constraint condition, we plot the fusion gain Q versus major radius R at four different confinement cases H=1.5, H=3 and H=5 and H=10 while ion temperature, normalized beta and wall loading are fixed. Figure 7 indicates the fusion gain is strongly dependent on major radius and H, both of which have a positive effect on Q by increasing the energy confinement time.

H=1.5 is the confinement enhancement factor that can be achieved by current experiments, and the minimum value of major radius which could meet Q=1 is 5m. In p-$^{11}$B reactions, the core ion temperature could be higher than 300keV, in which the neo-classical transport decreases significantly. Once a new method has been found to effectively suppress the anomalous transport, the confinement enhancement factor of H=3, 5, 10 might be possible in the future. In any case, H=3 is an overly optimistic but still foreseeable confinement enhancement factor, and could reach Q=1.5 at $R_0$=3m as shown in Figure 7.

### III.F.  The impurity effect

In the preceding discussion, the helium concentration has been ignored. In an actual reactor, helium as a production of p-$^{11}$B reaction, would decrease the fusion power output by diluting the fuel and increase bremsstrahlung radiation power, which would result in the termination of fusion reaction eventually.

The number of helium ions could be obtained from the continuity equation

$$\frac{N_{He}}{\tau_{He}} = \dot{N}_{He},$$

where $N_{He}$ is number of helium ions, $\dot{N}_{He}$ is the helium ion generation speed and $\tau_{He}$ is the helium confinement time.

The helium ion generation speed is obtained by the fusion power

$$\dot{N}_{He}(10^{20}/s) = \frac{3P_{fus}}{8.7 \times 16}.$$

And the core helium ion density is

$$n_{He0} = \dot{N}_{He}(1 + S_n)\tau_{He}/V_P.$$

If we assume a breakeven case, $n_{i0} = 6 \times 10^{20} m^{-3}, T_{i0} = 380 keV, P_{fus} = 5400 MW \cdot m^{-3}$, $\tau_E = 5s$ and $\tau_{He} = 10\tau_E \approx 50s$, we could get $n_{He0} = 9.5 \times 10^{20} m^{-3}$, even higher than the fuel ion density. Since the helium ash is poisonous to the fusion performance, excessive density of helium ions should be prevented.

In figure 8, we plot Q versus the ratio of helium density to ion density at three different confinement cases H=3 and H=5 and H=10 while ion temperature, normalized beta and wall loading are set constant. The dependence of fusion gain on $f_{He}$ is illustrated in figure 8, which indicates $f_{He}$ does play an important role in fusion gain. At H=3 case, the highest ratio of $f_{He}$ that the device could tolerate is 5%, which means the divertor ash removal efficiency should be high enough to keep the helium confinement time below energy confinement time.

### III.G.    The effect of synchrotron radiation loss

In the previous discussion, all these optimistic predictions have been done without considering the synchrotron radiation loss, whose effect critical role can hardly be ignored in the relatively high strong magnetic field up to 10T. As one of the most important energy loss mechanisms, its effect has to be taken into consideration in the actual reactor, and in this section, we will discuss the effect of synchrotron radiation loss on p-$^{11}$B fuel.

The equation (30) gives the volume-integrated synchrotron radiation power loss, which is determined by averaged density, toroidal magnetic field on axis, effective electron temperature, wall reflectivity, effective minor radius and approximated plasma volume. The range of parameters such as electron temperature, electron density, toroidal magnetic field and the device size in p-$^{11}$B tokamak have been already discussed in the previous section, hence one possible way to reduce the synchrotron power loss density is to increase the wall reflectivity high enough to reflect most of the radiation back and get re-absorbed in the core plasma.

To give a better insight into the effect of the synchrotron radiation loss on p-$^{11}$B fusion, the synchrotron radiation loss has been considered in the model predictions while other assumptions and constraints are still the same, and the helium concentration is set as 5%. The change of fusion gain and ratio of synchrotron radiation to bremsstrahlung radiation power loss in 3 different high-confinement cases H=5, H=10 and H=20 with wall reflectivity varied from 0.5 to 0.99 has been shown in figure 9.

In figure 9, the fusion gain of three high-confinement cases is even less than 0.5 with a low wall reflectivity of 0.5. When the wall reflectivity increases to 90%, the fusion gain of three high-confinement cases cannot meet the breakeven condition, and the synchrotron radiation power loss is slightly higher than the bremsstrahlung power loss. When the wall reflectivity is greater than 0.96, the breakeven condition can be obtained in the high confinement case of H=20. However, in this case the wall reflectivity and confinement enhancement factor are both unrealistically high for the existing technology. The results reveal that the fusion gain of p-$^{11}$B fuel is strong affected by synchrotron radiation loss, and one of the biggest challenges to the p-$^{11}$B fusion reactor is the reduction of synchrotron radiation loss.

### III.H.   The parameters of future device design

In the actual reaction, helium, which is poisonous to fusion performance, and synchrotron radiation loss, which greatly decreases the fusion gain, should be taken into consideration. Based on the calculations of system code, five representative sets of p-$^{11}$B tokamak device parameters which set helium concentration to 5% have been given in Table II. Among these sets, the wall reflectivity is assumed as 1 in set A to D, and in set E the wall reflectivity is assumed as 0.95 as a comparison. To keep the Helium concentration ratio in an acceptable range, the helium confinement time is assumed to be less than the energy confinement time, which also could be seen in Table II.

A normal-aspect-ratio tokamak with A=2.5 in set A could be found, in which a high magnetic field of 36.8T is needed, indicating low-aspect-ratio tokamak would be a better choice for p-$^{11}$B plasma with extremely high pressure. Besides, the plasma-stored energy is much higher in the low-aspect-ratio tokamak from set B to D than that in set A with the same size of major radius.

Three low-aspect-ratio tokamaks with A=1.4 from set B to set D are shown in Table II. Compared with Set B and Set C, when one increases the ratio $f_T$ from 2 to 2.5, the requirement for confinement condition is reduced to meet the breakeven condition. When one considers a hypothetical confinement enhancement factor H=10 in set D, fusion gain Q=4.14 could be obtained, which indicates a more optimistic confinement condition would bring a greater economic benefit. In set E, the effect of synchrotron radiation loss is considered by changing the wall reflectivity from 1 to 0.95, and the fusion gain is decreased from 4.14 to 0.84, which shows the synchrotron radiation loss would greatly decrease the fusion gain.

# IV. Conclusion and discussion

p-$^{11}$B reaction is hard to meet Lawson criterion because of its low reactivity. However, if we choose the optimum proton-Boron mix and ion temperature, and assume the ratio of ion to electron temperature to be 2.5, the fusion power produced by fuel ions of p-$^{11}$B can be comparable to heating power meeting Lawson criterion with Q=1.

The helium is poisonous to the fusion performance by diluting the fuel and increasing bremsstrahlung radiation loss, and excessive density of helium ions would result in the termination of fusion reaction. In order to keep Helium concentration ratio in acceptable range, the technique of active ash removal is needed to develop in order to reduce the helium confinement time less than energy confinement time.

Based on the calculation of tokamak system code, the fusion gain increases with the increasing major radius and confinement enhancement factor at fixed wall loading limit and beta limit. A major radius of 5m is needed to obtain fusion gain Q=1 with the current confinement condition of H=1.5 without helium concentration. After considering the helium concentration, if one can achieve a more optimistic confinement enhancement factor of H=10, the fusion gain of Q=4.14 at $R_0$=3m could be found, which is of economic benefit.

Although the ion temperature of above 300keV and energy confinement time of 10s required for p-$^{11}$B reaction is not achievable by existing technologies, the main point of this paper is to study the requirements needed for p-$^{11}$B fusion reactor, not to give engineering solutions to the requirements. In this point, this paper could achieve its aim if calls the colleagues' attention to the p-$^{11}$B fusion.

Finally, we should bear in mind that all these optimistic assumptions have been done by neglecting the synchrotron radiation loss, which is relatively high in the strong magnetic field. If we consider synchrotron radiation loss and assume a high wall reflectivity of 95% in the calculations of the case H=10, the fusion gain would decrease from 4.14 to 0.84. The results shows the p-$^{11}$B fusion reactor will not come true unless some techniques have been found in the future to avoid excessive synchrotron radiation loss.

## Acknowledgement

This work is supported by the China central government which guides the development of local science and technology funding No. 206Z4501G and the Compact Fusion Project of the ENN group.

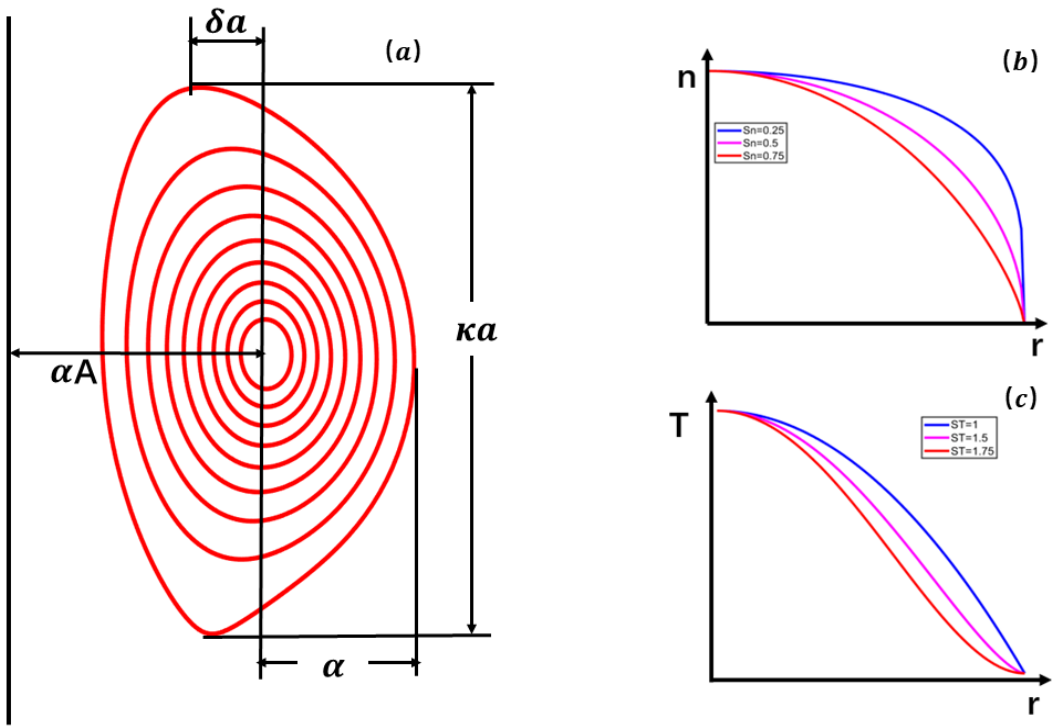

**Fig. 1.** The left figure is the sketch of a low -aspect -ratio tokamak geometry (a), and the right figure is a broader profile of density (b) and a steeper profile of temperature (c). In this paper, $S_n$=0.5 and $S_T$=1.5.

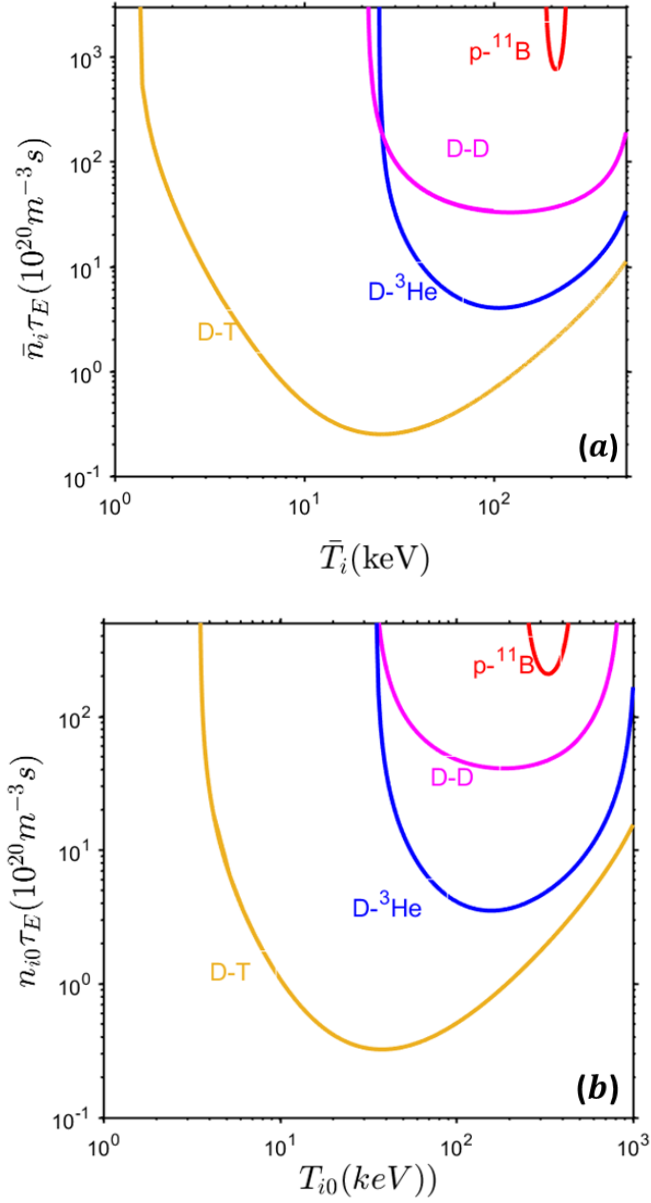

**Fig. 2.** Upper panel (a) is $\bar{n}_i \tau_E$ versus volume averaged ion temperature $\bar{T}_i$ which meet Lawson criterion for D-T, D-$^3$He, D-D and p-$^{11}$B reactions with Q=1 in the 0-dimensional model. Lower panel (b) is $n_{i0}\tau_E$ versus core ion temperature $T_{i0}$ which meet Lawson criterion for D-T, D-$^3$He, D-D and p-$^{11}$B reactions with Q=1 in the 1-dimensional model. In both models, $T_i = T_e$ for D-T, D-$^3$He and D-D reactions. As for p-$^{11}$B reaction, $\frac{n_p}{n_B} = 9$ and $\frac{T_i}{T_e} = 1$ in 0-dimensional model, while $\frac{n_p}{n_B} = 9$ and $\frac{T_i}{T_e} = 1.2$ in 1-dimensional model.

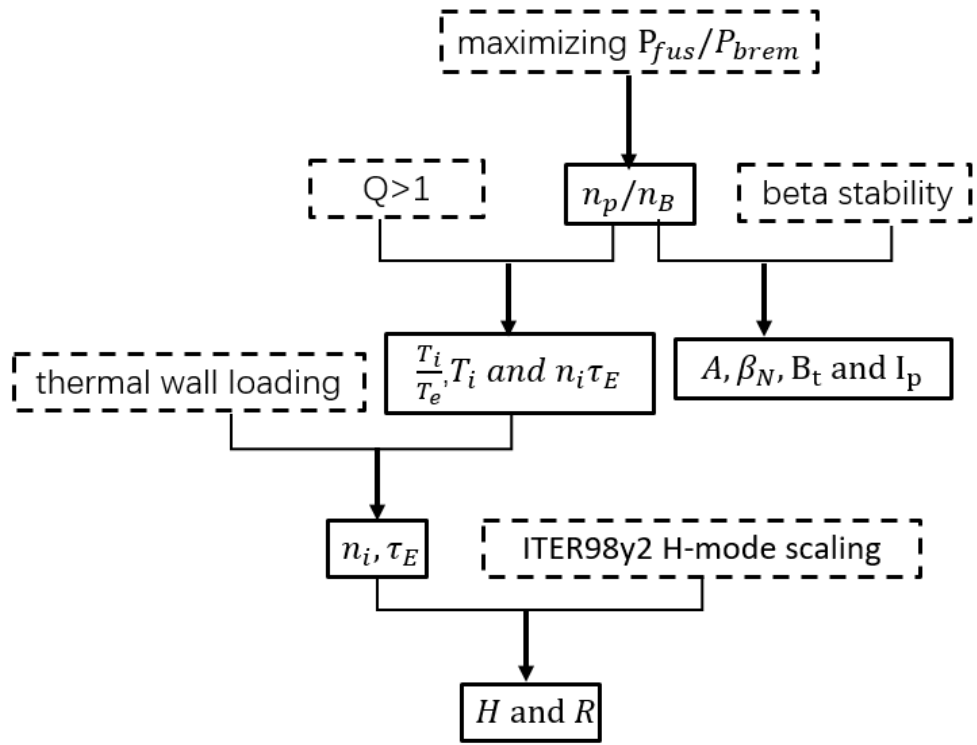

**Fig. 3.** The logic diagram of requirements for p-[11]B fusion reactor.

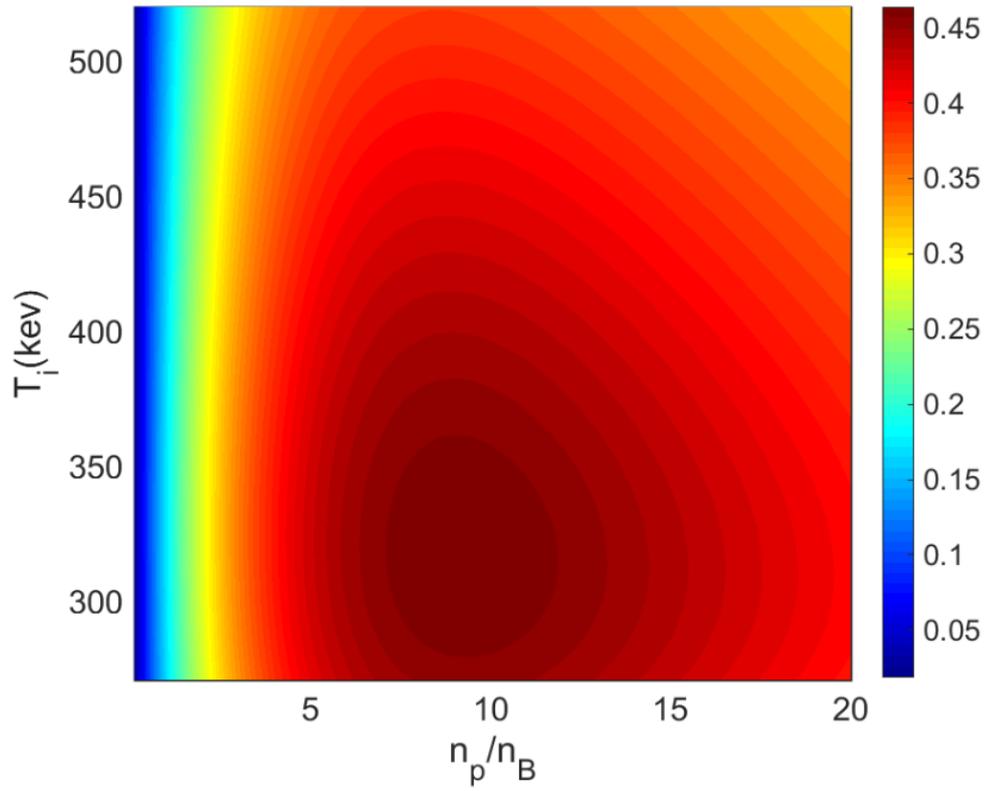

**Fig. 4.** The color map of $\frac{P_{fus}}{P_{brem}}$ versus $T_i$ and $f_{pB}$ for the constant $f_T = 1$. To maximize output of fusion power, the optimal value of $\frac{n_p}{n_B}$ is chosen as 9.

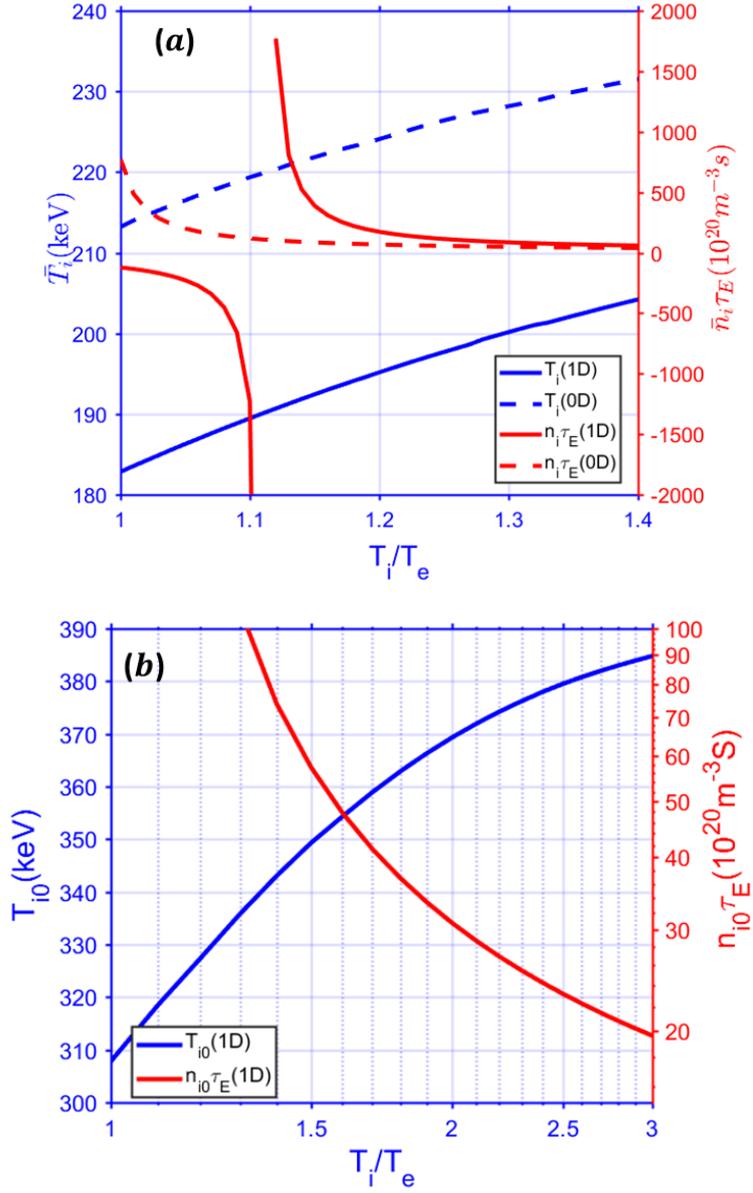

**Fig. 5.** The upper figure (a) is the minimum value of $\bar{n}_i \tau_E$ which meets Lawson criterion and the corresponding volume averaged ion temperature $\bar{T}_i$ versus $T_i/T_e$ in zero-dimensional model and in one-dimensional model. When $T_i/T_e < 1.12$, no positive solution that meet Lawson criterion with Q=1 can be found. The lower figure (b) is the minimum value of $n_{i0} \tau_E$ which meets Lawson criterion and the corresponding core ion temperature $T_{i0}$ versus $T_i/T_e$ in one-dimensional model.

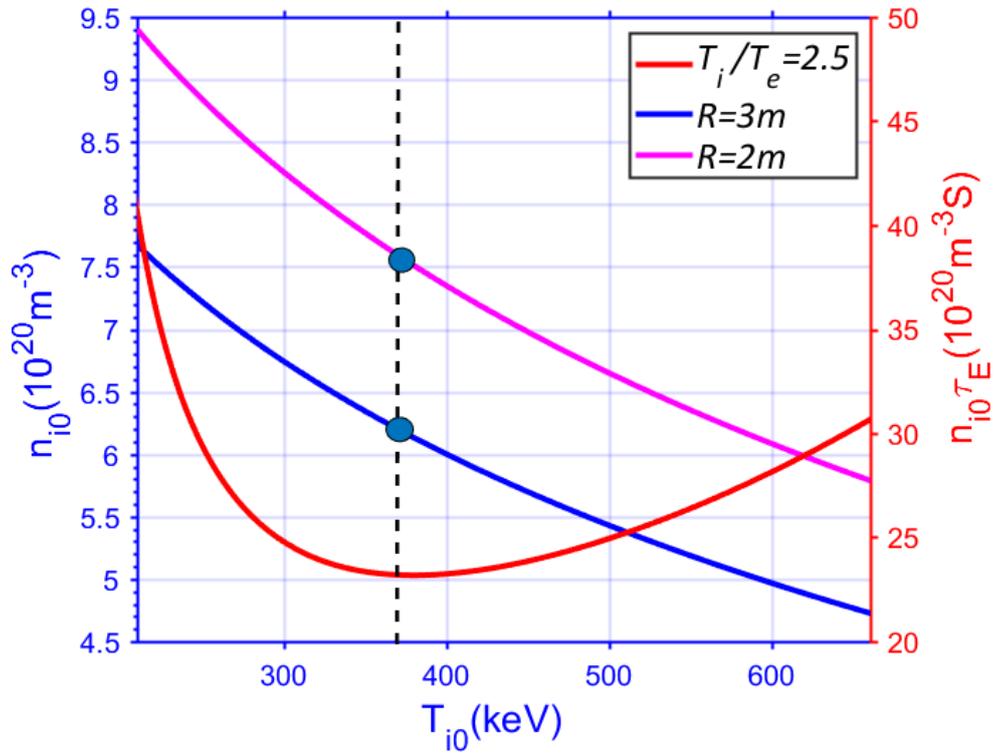

**Fig. 6.** The left Y axis shows the ion density versus ion temperature for bremsstrahlung radiation power density limit, and the right Y axis shows $n_i \tau_E$ versus ion temperature meeting Lawson criterion with Q=1. The magenta and blue line is the wall loading limit by radiation when R=2m and R=3m, and the circles intersected by optimum temperature of $T_i = 380 keV$ and the curve of bremsstrahlung radiation power limit indicate the marginal ion density.

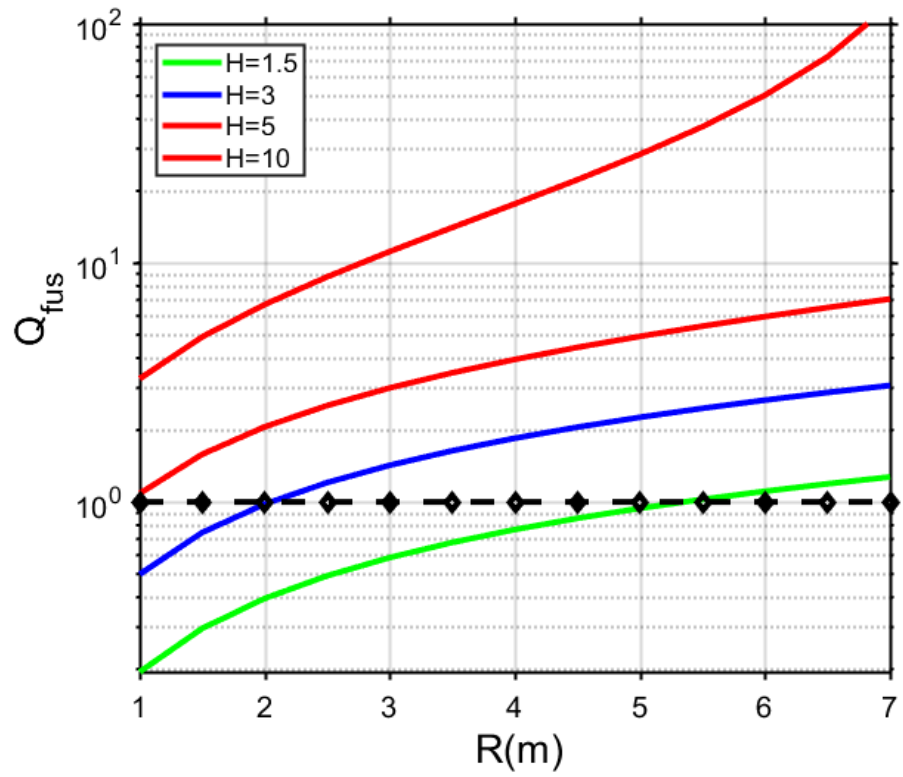

**Fig. 7.** Fusion gain Q versus major radius R at four different confinement cases H=1.5, H=3, H=5 and H=10.

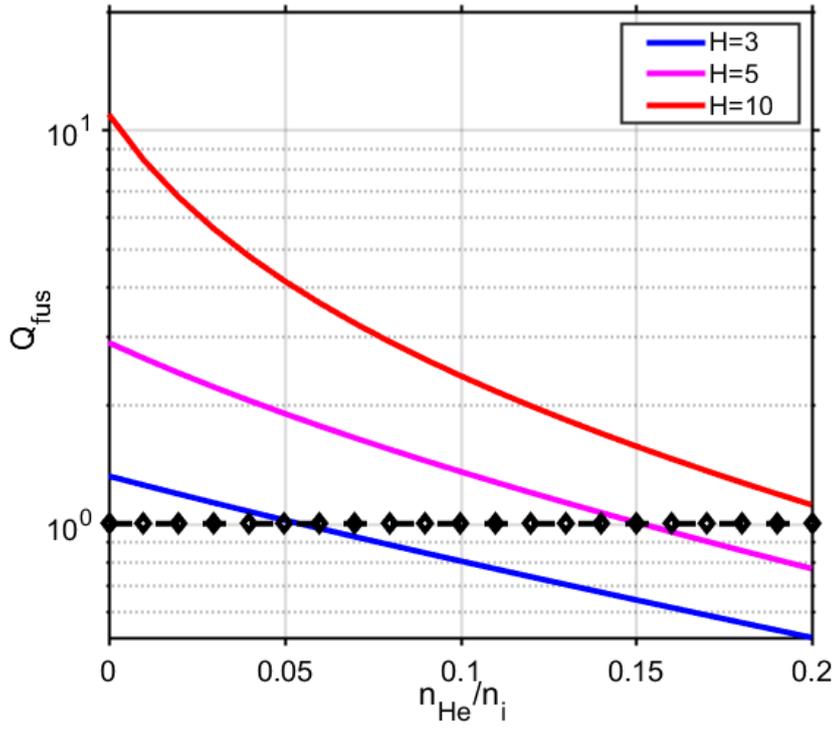

**Fig. 8.** Fusion gain Q versus the ratio of helium density to ion density at three different confinement cases H=3, H=5 and H=10.

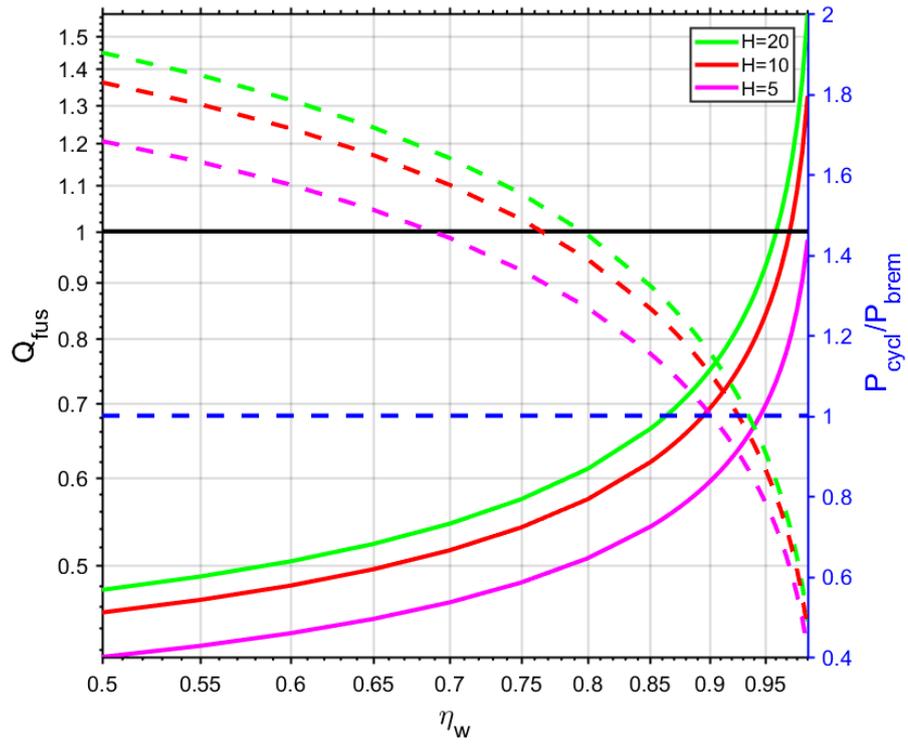

**Fig. 9.** Fusion gain (left y-axis) and the ratio of synchrotron radiation to bremsstrahlung radiation power loss (right y-axis) versus wall reflectivity at three different confinement cases H=5, H=10 and H=20.

**TABLE I**. The aspect ratio versus the minimum value of magnetic field required for confining p-$^{11}$B plasma and the corresponding plasma current.

| A | 1.4 | 1.9 | 2.4 |
|---|---|---|---|
| $B_{t0,min}$ (T) | 10.7 | 17.3 | 23.2 |
| Ip (MA) | 147 | 91.2 | 68.2 |

**TABLE II.** Five typical sets of p-$^{11}$B tokamak device parameter based on tokamak system code.

| Parameters | Set A | Set B | Set C | Set D | Set E |
|---|---|---|---|---|---|
| A | 2.5 | 1.4 | 1.4 | 1.4 | 1.4 |
| $f_T(T_i/T_e)$ | 2.5 | 2 | 2.5 | 2.5 | 2.5 |
| H | 3 | 5 | 3 | 10 | 10 |
| $f_{He}$ | 0.05 | 0.05 | 0.05 | 0.05 | 0.05 |
| $\eta_w$ | 1 | 1 | 1 | 1 | 0.95 |
| $R_0$(m) | 3 | 3 | 3 | 3 | 3 |
| a(m) | 1.2 | 2.14 | 2.14 | 2.14 | 2.14 |
| $V_p$(m$^3$) | 185 | 919 | 919 | 919 | 919 |
| $S_w$(m$^2$) | 233.3 | 548 | 548 | 548 | 548 |
| κ | 2.24 | 3.57 | 3.57 | 3.57 | 3.57 |
| δ | 0.5 | 0.5 | 0.5 | 0.5 | 0.5 |
| Q | 0.8 | 1.17 | 1.06 | 4.14 | 0.84 |
| $P_{fus}$(MW) | 2311 | 4424 | 5427 | 5427 | 5427 |
| $P_{aux}$(MW) | 2878 | 3771 | 5316 | 1311 | 6426 |
| $P_{cd}$(MW) | 576 | 754 | 1063 | 262 | 1292 |
| $P_{brem}$(MW) | 2333 | 5480 | 5480 | 5480 | 5480 |
| $P_{cycl}$(MW) | 0 | 0 | 0 | 0 | 4762 |
| $\phi_w$ (MW/m$^2$) | 10 | 10 | 10 | 10 | 10 |
| $B_{t0}$(T) | 36.8 | 11.3 | 11.4 | 12.9 | 11 |
| $I_p$(MA) | 63 | 140 | 140 | 124 | 145 |
| $W_{dia}$(MJ) | 7685 | 25864 | 26256 | 26256 | 26256 |
| $f_{bs}$ | 0.77 | 0.72 | 0.72 | 0.92 | 0.67 |
| $Z_{eff}$ | 2.4 | 2.4 | 2.4 | 2.4 | 2.4 |
| $T_{e0}$(keV) | 152 | 190 | 152 | 152 | 152 |
| $T_{i0}$(keV) | 380 | 380 | 380 | 380 | 380 |
| $n_{e0}$(10$^{20}$m$^{-3}$) | 12.4 | 7.72 | 8.55 | 8.55 | 8.55 |
| $n_{i0}$(10$^{20}$m$^{-3}$) | 8.70 | 5.40 | 5.98 | 5.98 | 5.98 |
| $\tau_E$(s) | 2.69 | 9.52 | 4.88 | 20.8 | 15.9 |
| $\tau_{He}$(s) | 1.08 | 1.73 | 1.57 | 1.57 | 1.57 |
| $\beta_T$ (%) | 5.12 | 36.7 | 36.8 | 28.8 | 39.5 |
| $\beta_p$ | 1.90 | 1.9 | 1.9 | 2.4 | 1.8 |
| $\beta_N$ | 3.6 | 6.4 | 6.4 | 6.4 | 6.4 |

| $q_{eng}$ | 3.14 | 2.22 | 2.2 | 2.8 | 2.06 |